\documentclass{ws-ijmpa}

\usepackage[super,compress]{cite}

\begin{document}
\markboth{N.V. Antonov, N.M. Gulitskiy, P.I. Kakin}{Dimensional transmutation and nonconventional scaling behaviour}

%
\catchline{}{}{}{}{}
%

\title{Dimensional transmutation and nonconventional scaling behaviour in a model of self-organized criticality}

\author{N.V. Antonov, N.M. Gulitskiy, P.I. Kakin, M.N. Semeikin}

\address{Department of Physics,  
Saint-Petersburg State University, 7/9~Universitetskaya nab. \\ 
St. Petersburg 199034, Russia\\
n.antonov@spbu.ru, n.gulitskiy@spbu.ru, p.kakin@spbu.ru}

\maketitle


\begin{abstract} 
{The paper addresses two unusual scaling regimes (types of critical behaviour) predicted by the field-theoretic renormalization group analysis for a self-organized critical system with turbulent motion of the environment. The system is modelled by the anisotropic stochastic equation for a ``running sandpile'' introduced by Hwa and Kardar in [{\it Phys. Rev. Lett.} {\bf 62}: 1813 (1989)]. The turbulent motion is described by the isotropic Kazantsev-Kraichnan ``rapid-change'' velocity ensemble for an incompressible fluid. The original Hwa-Kardar equation allows for independent scaling of the spatial coordinates $x_{\parallel}$ (the coordinate along the preferred dimension) and ${\bf x_{\bot}}$ (the coordinates in the orthogonal subspace to the preferred direction) that becomes impossible once the isotropic velocity ensemble is coupled to the equation. However, it is found that one of the regimes of the system's critical behaviour (the one where the isotropic turbulent motion is irrelevant) recovers the anisotropic scaling through ``dimensional transmutation.'' The latter manifests as a dimensionless ratio acquiring nontrivial canonical dimension. The critical regime  where both the velocity ensemble and the nonlinearity of the Hwa-Kardar equation are relevant simultaneously is also characterized by ``atypical'' scaling. While the ordinary scaling with fixed IR irrelevant parameters is impossible in this regime, the ``restricted'' scaling where the times, the coordinates, and the dimensionless ratio are scaled becomes possible. This result brings to mind scaling hypotheses modifications (Stell's weak scaling or Fisher's generalized scaling) for systems with significantly different characteristic scales.}

\end{abstract}

\section{Introduction \label{sec:Int}}

{The Hwa-Kardar differential stochastic equation\cite{HK,HK1} is a continuous model of self-organized criticality. Unlike equilibrium systems that require ``fine-tuning'' of control parameters (e.g., critical temperature in second order phase transitions) to become critical\cite{Book3}, systems with self-organized criticality arrive at critical state in the natural course of their evolution\cite{Bak3}. Such systems are wide-spread\cite{Col0,Col1,Col2,Col3} with examples ranging from neural systems to online social networks with many other systems in between.} 

{The Hwa-Kardar model is expected to describe 
the infrared (IR), i.e. large-scale and long-time asymptotic behaviour of the system with self-organized criticality and to yield the power laws that characterize the behaviour.}

{The system is an anisotropic ``running'' sandpile with an average flat profile that has a constant tilt along the preferred direction. The new sand keeps entering the system resulting in avalanches that cause other sand to exit the system. Eventually, the system arrives at a steady state with a self-similar profile that is anisotropic (due to the tilt).}

{The height of the profile is described by the scalar field $h(x)=h(t,{\bf x})$ where $t$ and a $d$-dimensional ${\bf x}$ stand for the time-space coordinates. Note that $h(x)$ is measured from the average tilted profile, i.e. it is a deviation of the profile height from its average value. The tilt corresponds to the preferred direction defined by the unit constant vector $ {\bf n}$. Any vector ${\bf q}$ can be decomposed as ${\bf q} = {\bf q}_{\perp} + {\bf n}\, q_{\parallel}$ where $({\bf q}_{\perp} \cdot \, {\bf n}) =0$. The spatial derivative ${\bf \partial}=\partial/ \partial {x_{i}}$, $i=1,\dots, d$, is then ``split'' into two derivatives: a derivative ${\bf \partial_{\perp}}=\partial/ \partial {x_{i}}$, $i=1,\,\dots,\, (d-1)$, in the subspace orthogonal to the vector $ {\bf n}$, and a one-dimensional derivative
$\partial_{\parallel} = ({\bf n} \cdot {\bf \partial})$.}

The stochastic differential equation that describes the evolution of the system is
\begin{equation}
\partial_{t} h= \nu_{\perp 0}\, {\bf \partial}_{\perp}^{2} h + \nu_{\parallel 0}\,
\partial_{\parallel}^{2} h -
\partial_{\parallel} h^{2}/2 + f.
\label{eq1}
\end{equation}
Here it is assumed that $\partial_{t} = \partial/ \partial {t}$ and ${\bf \partial}_{\perp}^{2}=({\bf \partial}_{\perp}\cdot{\bf \partial}_{\perp})$. The factors $\nu_{\parallel 0}$ and
$\nu_{\perp 0}$ stand for the diffusivity coefficients, and $f(x)$ is a Gaussian random noise with a zero mean and the correlation function
\begin{equation}
\langle f(x)f(x') \rangle = D_{0}\,
\delta(t-t')\, \delta^{(d)}({\bf x}-{\bf x}'), \quad D_0> 0.
\label{forceD}
\end{equation}

{Critical behaviour of a system is sensitive to various external disturbances including, in particular, motion of the environment\cite{Onuki,Nelson,Satten}. Thus, it is desirable to consider the effect of the motion on a system with self-organized criticality. }

{The coupling with the velocity field ${\bf v}(x)$ is achieved through a ``minimal substitution''
\mbox{$\partial_{t} \to \nabla_{t} = \partial_{t} + ({\bf v}\cdot {\bf \partial})$}.}

{Let us ascribe the following statistic to the velocity field ${\bf v}(x)$:
\begin{equation}
\begin{aligned}
\langle v_{i} (t, {\bf x}) v_{j}(t',{\bf x}')\rangle =  \delta(t-t')
D_{ij}({\bf x}-{\bf x}'),
\\
D_{ij}({\bf r}) = B_{0}
\int_{k>m} \frac{d{\bf k}}{(2\pi)^{d}} \frac{1}{k^{d+\xi}}
 P_{ij}({\bf k})
\exp ({\rm i} {\bf k}\cdot{\bf r}), \quad B_{0}>0.
\label{white}
\end{aligned}
\end{equation}
{This is the Kazantsev-Kraichnan ``rapid-change'' velocity ensemble\cite{FGV}.
Here $k\equiv |{\bf k}|$ stands for the wave number while $P_{ij}({\bf k})$ is
the transverse
projector defined in a standard way: $P_{ij}({\bf k}) = \delta_{ij} - k_i k_j / k^2$}. The velocity field is, thus, incompressible, i.e.
$({\bf  \partial}\cdot {\bf v}) = 0$. The IR regularization is provided by the sharp cutoff $k>m$. The parameter $\xi$ is non-negative and smaller than $2$.}

{In this paper, we apply field-theoretic renormalization group (RG) analysis to problem~(\ref{eq1})~--~(\ref{white}) and find out that it is the coupling of the anisotropic equation~(\ref{eq1}) and the isotropic velocity ensemble~(\ref{white}) that causes emergence of ``nonconventional'' scaling behaviour. In RG analysis, stochastic problem~(\ref{eq1})~--~(\ref{white}) is substituted with a field theory. Critical exponents from the power laws that describe the IR asymptotic behaviour of the system can be found from RG equations for the field theory. In particular, each IR attractive fixed point of 
the RG equations is connected to a regime of asymptotic behaviour, i.e. to a universality class. It was shown in [\refcite{WeU}] that the system's critical behaviour is divided into four universality classes. Two of them correspond to ordinary isotropic scaling in which IR irrelevant parameters are kept fixed and the coordinates $x_{\parallel}$ and ${\bf x_{\bot}}$ are not scaled independently. The other two regimes, however, involve nonconventional or anisotropic scaling. The aim of this paper is to explore these two regimes.}

\section{Renormalization and scaling of pure Hwa-Kardar model\label{sec:QFT1}}

{Let us begin by deriving scaling of the Hwa-Kardar equation without the turbulent motion of the environment. This ``pure'' scaling was first analysed by RG in its Wilsonion form in [\refcite{HK,HK1}]. We reproduce those results here (expressed in terms of the field theoretic RG) so that we can compare them with the scaling regimes of the model with turbulent advection.}

{Instead of the
stochastic problem~(\ref{eq1})~--~(\ref{forceD}), one can consider the equivalent field theory\cite{Book3} with the doubled set of fields $\Phi=\{h,h'\}$ and the action}
\begin{equation}
    S(\Phi) = \frac{1}{2}h'D_0\,h' + h'\left(-\partial_th + \nu_{\parallel0}\,\partial^2_{\parallel}h + \nu_{\perp0}\,{\bf \partial}^2_{\perp}h - \frac{1}{2}\,\partial_{\parallel}h^2\right).
    \label{action_pure}
\end{equation} 
{The integrations over the space-time coordinates $x=\{t,{\bf x}\}$ and the summations over the vector indices are implied throughout.}

{As model~(\ref{eq1}),~(\ref{forceD}) is anisotropic, it involves two independent spatial scales $L_{\parallel}$ and $L_{\perp}$ instead of a single scale $L$. Generally, a quantity's canonical dimension in dynamical models is described by the momentum dimension $d^{k}_{F}$ and the frequency dimension $d^{\omega}_{F}$ related to the spatial scale $L$ and the temporal scale $T$. Since there are two independent spatial scales, an arbitrary quantity $F$ is described by three canonical dimensions:}
\begin{equation}
    [F] \sim [T]^{-d^{\omega}_{F}}[L_{\parallel}]^{-d^{\parallel}_{F}}[L_{\perp}]^{-d^{\perp}_{F}}.
\end{equation}
{Table~\ref{canon_dim_pure} contains the canonical dimensions of all fields and parameters for the theory~(\ref{action_pure}) obtained from the condition that each term of the action~\eqref{action_pure} be dimensionless. We denoted a total canonical dimension $d_{F}$ as $d_{F} = d_{F}^{k} + 2d_{F}^{\omega}$ and a total momentum dimension $d_{F}^{k}$ as $d_{F}^{k} = d_F^{\parallel} + d_F^{\perp}$. The coupling constant $g_0$ is defined as} $D_0 = g_0\nu_{\perp0}^{3/2}\nu_{\parallel0}^{3/2}$.

{We can see from Table~\ref{canon_dim_pure} that  $g_0\sim\Lambda^{\varepsilon}$ where $\Lambda$ is a characteristic ultraviolet momentum scale and $\varepsilon= 4-d$, therefore, the logarithmic dimension is $d=4$. Let us also define the parameter $u_0 = \nu_{\parallel0}/\nu_{\perp0}$ that we will need later. Note that it possesses nontrivial momentum canonical dimensions $d_u^\parallel$ and $d_u^\perp$ while its total canonical dimension $d_{u}$ is zero.}

\begin{table}[t]
\tbl{Canonical dimensions for the theory~(\ref{action_pure}); $\varepsilon=4-d$.}
{\begin{tabular}{cccccccccc} \toprule
$\pmb F$ & $\pmb h'$ & $\pmb h$ & $\pmb D_0$ & $\pmb \nu_{\pmb \parallel\pmb 0}$ &$ \pmb \nu_{\pmb \perp\pmb 0}$& \boldmath {$ u_0$}&\boldmath {$g_0$}&$\pmb g$&$\pmb \mu$\pmb ,$\pmb m$\pmb ,$\pmb \Lambda$ \\  \colrule
$d_F^{\omega}$&$-1$&$1$&$3$&$1$&$1$&$0$&$0$&$0$&$0$\\ 
$d_F^{\parallel}$&$2$&$-1$&$-3$&$-2$&$0$&$-2$&$0$&$0$&$0$\\ 
$d_F^{\perp}$&$d-1$&$0$&$1-d$&$0$&$-2$&$2$&$\varepsilon$&$0$&$1$\\ 
$d_F^{k}$&$d+1$&$-1$&$-d-2$&$-2$&$-2$&$0$&$\varepsilon$&$0$&$1$\\ 
$d_F$&$d-1$&$1$&$4-d$&$0$&$0$&$0$&$\varepsilon$&$0$&$1$\\ \botrule
\end{tabular} \label{canon_dim_pure}}
\end{table}

{Canonical dimensions analysis augmented with symmetry analysis reveals that the model~(\ref{action_pure}) is multiplicatively renormalizable. It involves the sole nontrivial renormalization constant $Z_{\nu_{\parallel}} $. The only parameters that require renormalization are $\nu_{\parallel0}$ and $g_0$. They are related to their renormalized~counterparts in the following way ($\mu$ is the renormalization mass):}
\begin{equation}
\nu_{\parallel0} = \nu_{\parallel} Z_{\nu_{\parallel}} , \quad  
g_0= g\mu^{\varepsilon} Z_g.
\label{REN1}
\end{equation}

{The canonical differential equations for a renormalized Green function \mbox{$G^R=\langle \Phi \dots \Phi \rangle$} read}
\begin{equation}
\begin{aligned}
    \left(\sum_i d_i^{\omega} {\cal D}_i - d_{G}^{\omega} \right)G^R &= 0, \\
    \left(\sum_i d_i^{\perp} {\cal D}_i - d_{G}^{\perp} \right)G^R &= 0,\quad
    \left(\sum_i d_i^{\parallel} {\cal D}_i - d_{G}^{\parallel} \right)G^R &= 0.
\label{ScInvX}
\end{aligned}
\end{equation}
{The index $i$ enumerates the arguments of $G^R$ which are $\omega$, $k_{\perp}$, $k_{\parallel}$, $\mu$, $\nu_{\perp}$, and $\nu_{\parallel}$ (or $u$ depending on our choice of parameters). The operator ${\cal D}_a$ is ${\cal D}_a=a\partial_a$ for any $a$.}

{To calculate the critical exponents (or dimensions) that stand in scaling power laws, we need to supplement equations~(\ref{ScInvX}) with the differential RG equation}
\begin{equation}
{\left(
{\cal D}_{\mu} + \beta_g\partial_g - \gamma_{\nu_\parallel}{\cal D}_{\nu_\parallel}
- \gamma_{\nu_\perp}{\cal D}_{\nu_\perp} - \gamma_G \right) G^{R} = 0.}
\label{RGeqX}
\end{equation}
{Here $\gamma$ and $\beta$ are RG functions 
(anomalous dimensions and a  $\beta$-function respectively)}
{for the parameters of the system and coupling constant $g$. Note that quantities that are not renormalized have vanishing anomalous dimensions, i.e. $\gamma_{\nu_{\perp}}=0$ in the model~(\ref{action_pure}), but we keep this term for the future.}

{Universality classes of asymptotic (critical) behaviour are determined by fixed points of the RG equations, i.e. by the zeroes of the $\beta$ functions, $\beta(g^{*})=0$. Real parts of all the eigenvalues $\lambda_i$ of the matrix $\Omega_{ij}=\partial_{g_i}\beta_{g_j}$ (here $g =\{g_i\}$ is a full set of the charges) must be positive for a fixed point to be IR attractive.}

The substitution $g\to g^{*}$ and, hence, $\gamma_F \to\gamma_F ^*$ turns equation~\eqref{RGeqX} into equation with constant coefficients, i.e. into equation of the same type as equations~\eqref{ScInvX}.
Each of these equations describes a certain independent scaling behaviour of the function~$G^R$, in which some of its variables are scaled and some are kept fixed. A~parameter is scaled if the corresponding derivative enters the differential operator; otherwise the parameter is fixed.

We are interested in the critical scaling behaviour where the frequencies and momenta (or, equivalently, times and coordinates) are scaled, while the IR irrelevant parameters \mbox{(namely, $\mu$, $\nu_{\perp}$
and $\nu_{\parallel}$)} are kept fixed. 
Thus, we combine all these equations to exclude 
the derivatives with respect to all the IR irrelevant parameters
and arrive at the critical scaling equation for a given fixed point:
\begin{equation}
    \left({\cal D}_{k_\perp} +{\cal D}_{k_\parallel} \Delta_\parallel+ \Delta_\omega {\cal D}_\omega
        - \Delta_G  \right)G^{R} = 0
    \label{HK-scalingXZ}
\end{equation}
with $\Delta_{\parallel} = 1 + \gamma_{\nu_{\parallel}}^{*}/2$ and $\Delta_\omega=2-\gamma_{\nu_\perp}^*$. 

As we will see below, such an exclusion is not always possible: it requires some balance between the numbers of IR relevant and IR irrelevant parameters and the number of independent scaling equations.

The model~\eqref{action_pure} has two fixed points. The Gaussian (free) fixed point with $g^{*}=0$ is IR attractive for $\varepsilon<0$; the corresponding
critical dimensions coincide with canonical ones. 
The nontrivial fixed point with $g^{*}=32\varepsilon/9+O(\varepsilon^2)$ is IR attractive for $\varepsilon>0$. The corresponding canonical dimensions read
\begin{equation}
        \Delta_h = 1-\varepsilon/3, \quad \Delta_{h'} = 3 - \varepsilon/3, \quad 
        \Delta_{\omega}=2, \quad   \Delta_{\parallel} = 1+ \varepsilon/3.
        \label{HKcrit}
\end{equation}

\section{Renormalization and fixed points of the model with turbulent advection\label{sec:QFT2}}

As was stated above, inclusion of turbulent advection in the Hwa-Kardar model is achieved by replacing $\partial_t$ with $\nabla_t$ in the action~(\ref{action_pure}) and adding Gaussian averaging over the velocity ensemble with the correlation function~(\ref{white}). This leads to the following action functional: 
\begin{equation}
\begin{aligned}
    S(\Phi) = \frac{1}{2}h'D_0\,h' + h'\left(-\nabla_th + \nu_{\parallel0}\partial^2_{\parallel}h + \nu_{\perp0}\partial^2_{\perp}h - \frac{1}{2}\partial_{\parallel}h^2\right) + S_{{\bf v}},
    \label{action_ve}
\end{aligned}    
\end{equation} 
\begin{equation}
    S_{{\bf v}} = -\frac{1}{2}\int dt \int d{\bf x} \int d{\bf x}' v_i(t,{\bf x})D^{-1}_{ij}({\bf x}-{\bf x}')v_j(t,{\bf x}').
\label{action_Sv}
\end{equation}
Here $D^{-1}_{ij}({\bf x} - {\bf x'})$ is the kernel of the inverse operator $D^{-1}_{ij}$ for the integral operator $D_{ij}$ from~(\ref{white}) as $S_{{\bf v}}$ describes  Gaussian averaging over the field ${\bf v}$ with the correlation function~(\ref{white}).
  
Since that velocity ensemble is isotropic, it is no longer possible to define two independent spatial scales in this model. Thus, 
\begin{equation}
    [F] \sim [T]^{-d^{\omega}_{F}}[L]^{-d^{k}_{F}}
    \label{t5}
\end{equation}
and $d_{F} = d_{F}^{k} + 2d_{F}^{\omega}$.
Canonical dimensions of the fields and parameters of the model are presented in Table~\ref{canon_dim_ve}.
We note that for the quantities that 
are present in the pure Hwa-Kardar model with the action
(\ref{action_pure}) these dimensions coincide with their counterparts in Table~\ref{canon_dim_pure}.

\begin{table}[t]
\tbl{Canonical dimensions for the theory~(\ref{action_Sv}); $\varepsilon=4-d$.}
{\begin{tabular}{ccccccccccccc}
\toprule
\boldmath {$F$} & \boldmath {$h'$} & \boldmath {$h$} & \boldmath {$D_0$} & \boldmath {$\nu_{\parallel0}$} &\boldmath {$ \nu_{\perp0}$}&\boldmath {$g_0$}&\boldmath {$v$}&\boldmath {$B_0$}&\boldmath {$x_0$}&\boldmath {$u_0$},\boldmath {$u$}&\boldmath {$g$},\boldmath {$x$}&\boldmath {$\mu$}, \boldmath {$m$}, \boldmath {$\Lambda$} \\ \colrule
$d_F^{\omega}$&$-1$&$1$&$3$&$1$&$1$&$0$&$1$&$0$&$0$&$0$&$0$&$0$\\ 
$d_F^{k}$&$d+1$&$-1$&$-2-d$&$-2$&$-2$&$\varepsilon$&$-1$&$\xi-2$&$\xi$&$0$&$0$&$1$\\ 
$d_F$&$d-1$&$1$&$4-d$&$0$&$0$&$\varepsilon$&$1$&$\xi$&
$\xi$&$0$&$0$&$1$\\\botrule
\end{tabular}
\label{canon_dim_ve}}
\end{table}

In contrast to the pure Hwa-Kardar model and its modifications with anisotropic velocity ensembles\cite{AK1,SerovEP,Serov} (where the diffusivity coefficients $\nu_{\parallel0}$ and $\nu_{\perp0}$ have different momentum dimensions $d^\parallel$ and $d^\perp$), only total momentum dimension $d^k$ can be defined for the theory~\eqref{action_ve}. 
From Table~\ref{canon_dim_ve} it follows that
the ratio $u_0=\nu_{\parallel0}/\nu_{\perp0}$ is completely dimensionless, that is, dimensionless with respect to the frequency and momentum dimensions separately. 
Therefore, according to the general rules, $u_0$~should be treated as an additional charge.  
The third coupling constant is $x_0\sim\Lambda^{\xi}$ related to the amplitude $B_0$ of the correlation function~(\ref{white}) as
$B_0 = x_0\nu_{\perp0}$. The theory is logarithmic when $\varepsilon=0$ (i.e., $d=4$) and $\xi=0$. 

The theory is renormalized by introducing the renormalization constants $Z_i$:
\begin{equation}
\begin{aligned}
    \nu_{\parallel0} = \nu_{\parallel} Z_{\nu_{\parallel}} , \quad &\nu_{\perp0} = \nu_{\perp} Z_{\nu_{\perp}}, \quad 
    g_0=Z_g g\mu^{\varepsilon},\quad x_0=&Z_x x\mu^{\xi},\quad  u_0 = Z_u u.
    \label{renorm_param}
\end{aligned}    
\end{equation}
The anomalous dimensions calculated from the Feynman graphs have the following one-loop expressions (we omit details of calculations for brevity; see~[\refcite{AKL}] for similar graphs): 
\begin{equation}
    \gamma_{\nu_{\parallel}} = \frac{3}{8}\frac{x}{u} + \frac{3}{16}g, \quad \gamma_{\nu_{\perp}} = \frac{3}{8}x.
    \label{gamma1_gamma2}
\end{equation}
The anomalous dimensions for the coupling constants read
\begin{equation}
    \gamma_{g} = - \frac{3}{2}\left(\gamma_{\nu_{\parallel}} + \gamma_{\nu_{\perp}}\right), \quad \gamma_x = -\gamma_{\nu_{\perp}}, \quad \gamma_u = \gamma_{\nu_{\parallel}} - \gamma_{\nu_{\perp}}.
    \label{an-2}
\end{equation}
Similarly to the case of the pure Hwa-Kardar model (see Section \ref{sec:QFT1}), the~anomalous dimensions for the fields $h$, $h'$, and ${\bf v}$ vanish due to the absence of their renormalization: \mbox{$\gamma_h=\gamma_{h'}=\gamma_v=0$}. 

From explicit results for the anomalous dimensions~(\ref{gamma1_gamma2}), it follows that the  one-loop expressions for the $\beta$ functions have the form
\begin{equation}
\begin{aligned}
\beta_g &= g\left(-\varepsilon +\frac{9}{32}g+\frac{9}{16}\frac{x}{u}+\frac{9}{16}x\right), \quad
\beta_x =x\left(-\xi+\frac{3}{8}x\right), \\
\beta_u &= u\left(-\frac{3}{16}g-\frac{3}{8}\frac{x}{u}+\frac{3}{8}x\right).
\end{aligned}
\label{beta-expr1}
\end{equation}
Analysis of the system~\eqref{beta-expr1}
reveals two possible IR attractive fixed points: the Gaussian point FP1 with the coordinates $g^{*}=0$, $x^*=0$ and arbitrary
$u^*$, and the fixed point FP2 with the coordinates $g^{*}=0$, $x^*=8\xi/3$, $u^*=1$ which corresponds to the regime of simple turbulent advection
(the nonlinearity of the Hwa-Kardar equation is irrelevant in the sense of Wilson). 

The point FP1 is IR attractive for $\varepsilon<0$, $\xi<0$, whilst
the point FP2 is IR attractive for $\xi>\varepsilon/3$, $\xi>0$. This is the full set of fixed points with finite and nonzero value of $u^*$; there are no fixed points with $g^{*}\neq 0$ in the set, i.e. the Hwa-Kardar universality class is not realized for such values of $u$. 

However,  the system~(\ref{beta-expr1}) may lose possible solutions with $u^*=0$ or $1/u^*=0$. 
In order to explore those exceptional values, we have to pass to new variables: $w=x/u$ (to study the case 
$u^*=0$) and $\alpha=1/u$ (to study the case 
$u^*\to\infty$).

For the first case we obtain no IR attractive fixed points. 
In the second case we have
two more fixed points:
the point FP3 with the coordinates $g^{*}=32\varepsilon/9$, $x^*=0$, $\alpha^*=0$ and the point FP4 with  $g^{*}=32\varepsilon/9-16\xi/3$, $x^*=8\xi/3$, $\alpha^*=0$. The point FP3 corresponds to the regime of critical behaviour where only the nonlinearity of the Hwa-Kardar equation is relevant, and the point FP4 corresponds to the regime where both the nonlinearity
and the turbulent advection are relevant. The point FP3 is IR attractive for $\varepsilon>0$, $\xi<0$, and the point FP4 is IR attractive for $\xi<\varepsilon/3$, $\xi>0$. 

\vspace{-.3em}
\section{Scaling Regimes and Critical Dimensions in the Model with Turbulent Advection\label{sec:scal}}

Unlike model~(\ref{action_Sv}), the original Hwa-Kardar model~\eqref{action_pure} allows to introduce two independent momentum dimensions. 
Thus, it involves three equations~(\ref{ScInvX}) related to the canonical scale invariance. On the other hand, the model with turbulent advection contains only two such equations.
Combining them with differential RG equation
\begin{equation}
\left(
{\cal D}_{\mu} + \beta_g\partial_g + \beta_x\partial_x+ \beta_u\partial_u - \gamma_{\nu_\perp}{\cal D}_{\nu_\perp} -\gamma_G\right) G^{R} = 0
\label{RGeqXX}
\end{equation}
taken at a fixed point, we arrive at
\begin{equation}
    \left({\cal D}_{k_\parallel}+{\cal D}_{k_\perp} +\Delta_\omega{\cal D}_\omega-d^k_G -\Delta_\omega d^\omega_G-\gamma^*_G\right)G^{R} = 0,
    \label{G-scaling}
\end{equation}
where $\Delta_\omega=2-\gamma_{\nu_\perp}^*$.
This is the critical scaling equation; note that it does not involve derivatives over $\mu$ and $\nu_\perp$. 
The critical dimension $\Delta_F$ of a quantity $F$ now reads
\vspace{-.9em}
\begin{equation}
    \Delta_F=d^k_F +\Delta_\omega d^\omega_F+\gamma^*_F.
    \label{crete}
\end{equation}
For the fixed points FP2 critical dimensions are 
\begin{equation}
    \Delta_h=1-\xi, \quad \Delta_v=1-\xi, \quad \Delta_{h'}=3-\varepsilon+\xi, \quad \Delta_{\omega}=2-\xi.
\end{equation}
These results are perturbatively exact due to the fact that $\gamma_{\nu_\perp}^*$ is known exactl while $\gamma_{h}=\gamma_{h'}=\gamma_{v}=0$. 

The fixed point FP1 is Gaussian so the corresponding critical dimensions coincide with the canonical ones. 

Surprisingly, the fixed points FP3 and FP4 are related to scaling that is majorly different from the scaling described by the points FP1 and FP2. The difference arises when one sets $\beta_{\left\{g_i^*\right\}}=0$ in the RG equation~(\ref{RGeqXX}). The Green functions have well-defined finite limits at $g\to0$ and $x\to 0$, thus, the $\beta$-functions at the points FP1 and FP2 can be set to zero without the need for any further analysis. However, such a straightforward substitution cannot be performed for $\beta_\alpha$ at the fixed points with $\alpha^*=0$ (or alternatively at the fixed points with $u\to\infty$). Instead,
the first nontrivial order of the expansion of $\beta_\alpha$ 
around such exceptional fixed points
should be retained in~(\ref{RGeqXX}). More detailed discussion of this important issue is given in Appendix A.
As a result, the critical scaling equation at the fixed points FP3 and FP4 takes on the form:
\begin{equation}
    \left({\cal D}_{k_\parallel}+{\cal D}_{k_\perp} +\Delta_\omega{\cal D}_\omega-
    \lambda^*{\cal D}_\alpha -d^k_G -\Delta_\omega d^\omega_G-\gamma^*_G\right)G^{R} = 0,
    \label{34-scaling}
\end{equation}
where $\lambda=\partial \beta_\alpha/\partial\alpha$ taken at $\alpha=0$, $\Delta_\omega=2-\gamma_{\nu_\perp}^*$, and $\lambda^*=\lambda(g^{*},x^*)$.

The point FP3 corresponds to the regime where only the nonlinearity of the Hwa-Kardar equation is relevant; its scaling equation takes the form
\begin{equation}
    \left({\cal D}_{k_\parallel}+{\cal D}_{k_\perp} +2{\cal D}_\omega-
    \frac{2\varepsilon}{3}
    \, {\cal D}_\alpha -d^k_G -2d^\omega_G-\gamma^*_G\right)G^{R} = 0.
    \label{3-scaling}
\end{equation}
This equation corresponds to scaling where the momenta $k_\parallel$, ${\bf k}_{\bot}$, the~frequency~$\omega$, and the ratio $\alpha$ are scaled. However, in this special case where only the nonlinearity is relevant, the model~\eqref{action_ve} coincides with the pure Hwa-Kardar model~\eqref{action_pure}. It means that additional canonical symmetry arises and turns $d^\parallel$ and~$d^\perp$ into independent dimensions, see~equations~\eqref{ScInvX}.

The general solution of the system of equations that includes two canonical invariance equations, the
RG equation~(\ref{RGeqXX}) taken at a fixed point, and~homogeneous counterpart of equation~\eqref{3-scaling} is an arbitrary function of three independent variables chosen here for definiteness as
\begin{equation}
    z_1=\frac{\omega}{\nu_\perp k_\perp^2}, \quad 
    z_2=\frac{k_\parallel}{k_\perp}, \quad \text{and} 
    \quad z_3=  \alpha \left(\frac{k_\perp}{\mu}\right) ^{2\varepsilon/3}.
    \label{sol1}
\end{equation}

Additional symmetry requires the variables $z_1$, $z_2$, and $z_3$ to be dimensionless with respect to Table~\ref{canon_dim_pure}.
While the variables $z_2$ and $z_3$ do not satisfy this requirement, it is possible to introduce a new variable $z_0=z_2z_3^{-1/2}$ with needed canonical dimensions that serves as the second solution (along with $z_1$) of the homogeneous part of the equation~\eqref{HK-scalingXZ}:
\begin{equation}
    z_0= 
    \frac{k_\parallel }{k_\perp^{\Delta_\parallel} \alpha^{\wp}}\, \mu^{\varepsilon/3} 
    \quad \text{with} 
     \quad \wp=\frac{3}{2\varepsilon}\left(\Delta_\parallel-1\right),
    \label{sol2}
\end{equation}
where $\Delta_\parallel=1+\varepsilon/3$ is in agreement with expressions~\eqref{HKcrit}. This implies that the fixed point FP3 allows for the scaling where the coordinates $x_\parallel$ and ${\bf x}_{\bot}$ (or momenta $k_\parallel$ and ${\bf k}_{\bot}$) 
are scaled simultaneously with nontrivial $\Delta_\parallel\neq1$, while all the IR irrelevant parameters (including $\alpha$) are kept fixed. Thus, the results derived for the pure Hwa-Kardar model~\eqref{action_pure} are reproduced.

The fixed point FP4 corresponds to the regime where both the nonlinearity of the Hwa-Kardar equation and the turbulent advection are relevant; its equation of critical scaling~\eqref{34-scaling} reads
\begin{equation}
    \left({\cal D}_{k_\parallel}+{\cal D}_{k_\perp} +\Delta_\omega{\cal D}_\omega-
    \left(\frac{2}{3}\varepsilon-2\xi\right){\cal D}_\alpha -d^k_G -\Delta_\omega d^\omega_G-\gamma^*_G\right)G^{R} = 0,
    \label{4-scaling}
\end{equation}
where $\Delta_\omega=2-\xi$. Three combinations $\hat{z}_1$, $\hat{z}_2$, and $\hat{z}_3$ are possible solutions of its homogeneous part:
\begin{equation}
    \hat{z}_1= \frac{\omega}{\nu_\perp k_\perp^{2}}\left(\frac{k_\perp}{\mu}\right)^\xi, \quad 
    \hat{z}_2= \frac{k_\parallel}{k_\perp}, \quad
    \hat{z}_3= \alpha \left(\frac{k_\perp}{\mu}\right) ^{2\varepsilon/3-2\xi}.
    \label{FP4-an}
\end{equation}
The variables~\eqref{FP4-an} describe generalized critical scaling where
$\alpha$ (i.e. the ratio of $\nu_\perp$ and~$\nu_\parallel$) is also scaled. 

This resembles various modified similarity hypotheses for systems with  different characteristic scales or different scaling laws\cite{Stell,Stell2,Stell3,Stell4}. 

Of course, it is not impossible that the Green functions  have some peculiar dependence on the parameters~(\ref{FP4-an}) so that the ordinary critical scaling at fixed $u$, with some nontrivial scaling dimensions, takes place. However, as far as we can see, there is no anticipated reason for this situation to occur.

\section{Conclusion \label{sect:Con}}

We studied nonconventional scaling behavior of the self-organized critical system in an insotropic turbulent environment. The system was modelled by the anisotropic Hwa-Kardar equation~(\ref{eq1})~--~(\ref{forceD}) while the turbulent advection was described by Kazantsev-Kraichnan ``rapid-change'' ensemble~(\ref{white}).  

We analyzed a field theory equivalent to the stochastic problem and established that the theory is multiplicatively renormalizable. The coordinates of the four fixed points of RG equations were calculated in the one-loop approximation or found exactly. 

Among four universality classes of critical behaviour, two classes turned out to correspond to nonconventional scaling behavior. It was shown that a kind of dimensional transmutation takes place in one of those regimes, precisely, in the one where the nonlinearity of the Hwa-Kardar equation is relevant. The transmutation results in the ratio $u$ of the two diffusivity coefficients $\nu_\parallel$ and $\nu_\perp$ acquiring a nontrivial canonical dimension. Due to this, the new canonical symmetry arises in the model~\eqref{action_ve} and allows canonical dimensions $d^\parallel$ and $d^\perp$ to be scaled independently. Thus, the scaling behaviour involves simultaneous scaling of the coordinates $x_\parallel$ and ${\bf x}_{\bot}$ with a nontrivial relative exponent $\Delta_\parallel\neq1$. The~regime is in agreement with the predictions of the pure Hwa-Kardar model analysis~\eqref{action_pure}.

In the regime where both the advection and nonlinearity are relevant, the self-similar behaviour necessarily involves some dilation of the ratio $u$. 
The ordinary critical behaviour with fixed definite critical dimensions can be established only if the correlation functions of the model have some special dependence on $u$. Otherwise, the IR behaviour will remind 
of various generalized self-similarity hypotheses, like the
weak scaling due to George Stell\cite{Stell,Stell2,Stell3}
or the parametric scaling in the spirit of Michael Fisher\cite{Stell4} 
for systems with different characteristic scales and modified scaling laws.

In the future, it would be interesting to consider advection by more realistic velocity ensembles and take into account finite correlation time, non-Gaussianity, anisotropy and so on. It is indeed a very interesting avenue to explore and sometimes introducing of ``small'' properties of turbulent motion leads to very unexpected resulting features, such as losing of universality~\cite{Serov} or logarithmic corrections to ordinary scaling (instead of power ones) in anisotropic advection of vector impuriy field\cite{Log,Log2,Log3,Log4}. However, it is impossible to predict the result from general considerations before calculations will be done.

This work is partly in progress. Preliminary investigation\footnote{The results were reported at the International Workshop on Statistical Physics (December 1-3, 2021; Antofagasta, Chile) by A.~Yu.~Luchin; see the poster ``A self-organized critical system under the influence of turbulent motion of the environment'' at  https://www.iwosp.cl/poster-session}
of the advection by the stochastic Navier-Stokes equation with  white stirring noise (model B of Ref.~[\refcite{FNS}])
has shown that the RG equation possess at least one IR attractive fixed point.  It corresponds to the regime of simple turbulent advection, where the nonlinearity of the Hwa-Kardar equation is irrelevant in the sense of Wilson and where isotropy is restored. While the system of $\beta$ functions is quite complicated even at one-loop level due to emergence of two new charges, existence of other fixed points seems to be a most likely possibility.

\section*{Acknowledgments}
The reported study was funded by the Russian Foundation for Basic Research, project number~20-32-70139.
The work by N.V.A.~and~ P.I.K. was also supported by the Theoretical Physics and Mathematics Advancement Foundation ``BASIS.''

\appendix
\section{Subtleties of RG equations}
\label{sec:Subs}

Let us illustrate the reasoning behind the derivation of equation~(\ref{34-scaling}) using the simplified model as an example; cf. the discussion of the magnetohydrodynamical turbulence in sec.~3.7 of the monograph~[\refcite{violet}].

Let $D=D(k,\mu,\nu,g)$ be the renormalized equal-time pair correlation function of a certain renormalizable dynamic model with the diffusivity coefficient $\nu$ and the coupling constant $g$. From the dimensionality considerations one can write:
\begin{equation}
D=k^{d^k_D+2d^{\omega}_D}\, \nu^{d^{\omega}_D}\, R(k/\mu,g),    \label{ex1}    
\end{equation}
where $d^k_D$ and $d^{\omega}_D$ are the canonical dimensions 
of $D$ and $R(\cdot)$ is a function of dimensionless arguments. 

The corresponding RG equation has the form
(here and below, see, e.g. the monograph~[\refcite{Book3}]):
\begin{equation}
\left({\cal D}_{\mu} + \beta(g)\partial_g-\gamma_{\nu}{\cal D}_{\nu} + \gamma_{D} \right) D=0,
\label{exRG}    
\end{equation}
and its solution is:
\begin{equation}
D=k^{d^k_D+2d^{\omega}_D}\, \bar\nu^{d^{\omega}_D}\, 
R(1,\bar g)\,
\exp \left\{ \int_g^{\bar g}\, \frac{\gamma_D(s)\,ds}{\beta(s)}
\right\},
\label{ex2}    
\end{equation}
where the functions  $\bar g=\bar g(k/\mu,g)$, $\bar \nu=\bar \nu(k/\mu,g,\nu)$ are the RG-invariant counterparts of the parameters  $g$, $\nu$ while $\gamma_{\nu}$, $\gamma_D$, $\beta$ are the RG functions.

If the RG equation has an IR attractive fixed point,
$\beta(g^*)=0$, $\omega=\beta'(g^*)>0$ (the case $g^*=0$ is allowed) in the IR asymptotic region $k/\mu\to 0$ one has:
\begin{equation}
\bar g - g^* \to (k/\mu)^{\omega}, 
\quad \bar \nu \to \nu\, (k/\mu)^{-\gamma_{\nu}^*}, \quad
\exp\{\cdots\} \to (k/\mu)^{\gamma_{D}^*}.
\label{ex3}    
\end{equation}
The IR asymptotic expression for $D$ is obtained by the  substitution of~(\ref{ex3}) into~(\ref{ex2}). 
If the function $R(1,g)$ has a finite limit at $g\to g^{*}$,
one can neglect the \mbox{term $\sim (k/\mu)^{\omega}$} in the expression for $\bar g - g^{*}$, decaying for
$k/\mu\to 0$, and simply substitute $\bar g \to g^{*}$.
This gives
\begin{equation}
D \simeq k^{\Delta_D}\, R(1,g^{*}),    
\label{ex4}    
\end{equation}
where the critical dimension ${\Delta_D}$ is given by the standard expression~(\ref{crete}) with $\Delta_{\nu}=2-\gamma_{\nu}^*$
(here and in similar expressions below we do not display the dependence on the fixed parameters $\mu$ and $\nu$). 

The expressions~(\ref{ex3}) and~(\ref{ex4}) can be more easily derived from the simplified RG equation with constant coefficients,
obtained from~(\ref{exRG}) by the replacement of the RG functions by their leading terms\footnote{Accounting for the higher-order corrections in $(g-g^{*})$ would give only $g$-dependent nonuniversal amplitudes in scaling laws and corrections to the leading-order IR asymptotic expressions for the correlation functions of the type~(\ref{ex4}); see the monograph~[\refcite{Book3}], Secs.~1.3~and~1.33.}
at $g\to g^{*}$, that is,
$\gamma_F(g)\to \gamma_F^*$, $\beta(g)\to \omega (g-g^{*})$:
\begin{equation}
   \left({\cal D}_{\mu} + \omega {\cal D}_{g-g^{*}}
   -\gamma_{\nu}^*{\cal D}_{\nu} + \gamma_{D}^* \right) D=0,
\label{exRG2}    
\end{equation}
where we used the identity $\partial /\partial g =
\partial /\partial (g-g^{*})$. Note that the asymptotic expressions for $\bar g$, $\bar \nu$ in ~(\ref{ex3}) are solutions of the homogeneous analog of equation~(\ref{exRG2}).

If the function $R(1,g)$ is finite  at $g= g^{*}$ and  
one can neglect the right hand side in the expression $\bar g - g^{*} \simeq (k/\mu)^{\omega}$ and the term coming from the $\beta$
function in~(\ref{exRG2}) can also be omitted; this is the usual situation.

If this is not the case, the IR behaviour requires a more careful analysis. Assume, as an example, that $R$ has the form
\begin{equation}
    R(k/\mu,g)= (g-g^{*})^a\,F(k/\mu,g),
\label{ex5}    
\end{equation}
where $a$ is some exponent and $F$ has a finite limit for $g\to g^*$. Then the naive substitution $\bar g\to g^*$ leads to the vanishing or divergence of the amplitude factor in~(\ref{ex4}). In order to find genuine IR behaviour, one should take into account the way in which the invariant coupling approaches its fixed point, namely: 
$\bar g - g^* \simeq  (k/\mu)^{\omega}$, where $\omega=\beta'(g^*)>0$.
This gives:
\begin{equation}
D \simeq k^{\Delta_D+a\omega}\, F(1,g^*).    
\label{ex6}    
\end{equation}
Thus, the correct critical dimension of the function $D$ appears to be $\Delta_D+a\omega$ rather than $\Delta_D$ itself.

In ordinary cases, expression ~(\ref{ex6}) determines one of the correction terms to the leading-order IR asymptotic expression~(\ref{ex4}); see Secs.~1.3 and~1.34 in the monograph~[\refcite{Book3}]. In our case, it becomes the leading-order term in itself. In order to derive it directly  from the simplified RG equation, one has to substitute~(\ref{ex1}) and~(\ref{ex5}) into~(\ref{exRG2}). This gives
\begin{equation}
   \left({\cal D}_{\mu} + \omega {\cal D}_{g-g^{*}}
   -\gamma_{\nu}^*{\cal D}_{\nu} + 
   \gamma_{D}^*+a\omega \right) F=0,
\label{exRG3}    
\end{equation}
where the desired replacement $\gamma_{D}^*\to \gamma_{D}^*+a\omega$ is due to the term with 
${\cal D}_{g-g^{*}}$ in~(\ref{exRG2}).
Now, since the function $F$ is finite at $g=g^{*}$, 
this term in the equation~(\ref{exRG3}) can be omitted,
which immediately leads to representation~(\ref{ex6}).
It is also worth noting that the functions $R$ and $F$ in~(\ref{ex5}) differ by the constant ($k$-independent) factor  $(g-g^{*})^a$ and not by the $k$-dependent factor $(\bar g-g^{*})^a$, so that their scaling behaviour is identical up to irrelevant nonuniversal amplitudes.

If the character of the singularity at $g=g^{*}$ is not known {\em a priori}, the operation ${\cal D}_{g-g^{*}}$ should be retained in the RG differential operator.
It is the situation that is encountered in Section~\ref{sec:scal} when analysing the fixed points FP3 and FP4 with the coordinate $\alpha^*=0$. That is, the term with the coefficient $\lambda^*=\beta_{\alpha}'(\alpha^*)$ in equation~(\ref{34-scaling}) must be retained.


\end{document}